\documentclass[acmlarge,screen]{acmart}
\usepackage{xcolor}
\usepackage{tabularx}
\usepackage{geometry}
\usepackage{titlesec}

\titleformat{\section}
  {\normalfont\Large\bfseries}{\thesection}{1em}{\MakeUppercase}
\titleformat*{\subsection}{\normalfont\Large\bfseries}

\AtBeginDocument{%
  \providecommand\BibTeX{{%
    \normalfont B\kern-0.5em{\scshape i\kern-0.25em b}\kern-0.8em\TeX}}}

\setcopyright{acmlicensed}
\acmJournal{PACMHCI}
\acmYear{2024} \acmVolume{8} \acmNumber{CSCW1} \acmArticle{20} \acmMonth{4} \acmPrice{15.00}\acmDOI{10.1145/3637297}

\begin{document}

\title{"Because Some Sighted People, They Don't Know What the Heck You're Talking About:" A Study of Blind TikTokers' Infrastructuring Work to Build Independence}


\author{Yao Lyu}

\orcid{0000-0003-3962-4868}
\affiliation{%
  \institution{Pennsylvania State University}
  \city{University Park}
  \state{Pennsylvania}
  \country{USA}
}
\email{yaolyu@psu.edu}

\author{John M. Carroll}
\orcid{0000-0001-5189-337X}
\affiliation{%
  \institution{Pennsylvania State University}
  \city{University Park}
  \state{Pennsylvania}
  \country{USA}
}
\email{jmcarroll@psu.edu}

\renewcommand{\shortauthors}{}


\begin{abstract}

There has been extensive research on the experiences of individuals with visual impairments on text- and image-based social media platforms, such as Facebook and Twitter. However, little is known about the experiences of visually impaired users on short-video platforms like TikTok. To bridge this gap, we conducted an interview study with 30 BlindTokers (the nickname of blind TikTokers). Our study aimed to explore the various activities of BlindTokers on TikTok, including everyday entertainment, professional development, and community engagement. The widespread usage of TikTok among participants demonstrated that they considered TikTok and its associated experiences as the infrastructure for their activities. Additionally, participants reported experiencing breakdowns in this infrastructure due to accessibility issues. They had to carry out infrastructuring work to resolve the breakdowns. Blind users' various practices on TikTok also foregrounded their perceptions of independence. We then discussed blind users' nuanced understanding of the \textit{TikTok-mediated independence}; we also critically examined BlindTokers' infrastructuring work for such independence.
\end{abstract}

\ccsdesc[500]{Human-centered computing~Human computer interaction (HCI)}
\ccsdesc[500]{Human-centered computing~Empirical studies in HCI}

\keywords{Visual Impairment, TikTok, BlindTok, BlindToker, Short-Video Platform, Infrastructuring, Infrastructure, Accessibility, Independence}

\maketitle

\section{Introduction}




Social media has become an indispensable aspect of human life. It offers opportunities for building friendships and exchanging ideas through various information and communication technologies (ICTs). However, individuals with visual impairments (PVI) encounter difficulties when it comes to consuming, creating, and sharing content on social media platforms due to inaccessible designs. Researchers have consistently focused on addressing the accessibility challenges faced by PVI on social media platforms \cite{Seo2018,10.1145/3313831.3376404}. They strive to find solutions that encompass a wide range of issues, from facilitating the consumption of image content \cite{10.1145/3313831.3376404} to promoting social interactions \cite{Wu2014}. Nonetheless, previous studies have predominantly concentrated on text- or image-based social media platforms such as Facebook and Twitter.


Recently, short-video platforms have gained rapid popularity among social media users \cite{bartolome2023literature}. As of January 2022, TikTok, the most popular short-video platform, boasts 1 billion monthly active users worldwide \cite{Geyser2022}. TikTok's immense popularity has also attracted a substantial number of visually impaired users. Although there is no official report on the exact number of blind users on TikTok, the author's search in September 2022 revealed approximately 5 million videos tagged with "BlindTok" (the nickname of the blind users' community on TikTok) or its variations published on the platform. This substantial number of videos related to "BlindTok\footnote{We use "BlindTokers", "TikTok users with visual impairments", and "blind TikTok users" interchangeably in this paper.}" suggests a significant population of blind individuals on TikTok and highlights the diverse and varied interactions between blind users and the platform. Notably, blind users' interactions with short-video platforms may differ significantly from those on platforms that focus on text or image because "\textit{video content... may impose additional burdens and challenges on people with visual impairments} (p.2) \cite{10.1145/3476038}." In fact, the HCI community has been paying growing attention to users with disabilities on TikTok. For instance, Duval et al. \cite{10.1145/3411764.3445303} presented a study on a TikTok user group that consisted of people with various disability statuses. The study reported how the users collaborated with each other to enjoy TikTok as a playful platform. However, limited research has been conducted to explore the motivations and experiences of BlindTokers on short-video platforms. In this study, we aim to investigate blind users' interactions with TikTok. Especially, we are interested in BlindTokers' motivations for using TikTok; we also want to study the challenges (especially challenges related to accessibility) they face and the countermeasures they come up with when using TikTok. We outline the research questions as follows:


\begin{enumerate}
  \item \textbf{\textit{What are BlindTokers' purposes of using TikTok?}}.
  \item \textbf{\textit{What are the challenges and solutions in terms of accessibility when BlindTokers use TikTok?}}
\end{enumerate}



We conducted an interview study to address the research question. We recruited 30 BlindTokers with diverse backgrounds and utilized semi-structured interviews to gather data. Subsequently, we conducted a thematic analysis to identify prominent patterns in their interactions with TikTok. The analysis revealed two key themes: TikTok as infrastructure for BlindTokers' multifaceted needs and TikTok's accessibility breakdowns and BlindTokers' work to overcome the breakdowns. We turned to concepts of infrastructure \cite{Star1996} and infrastructuring work \cite{Pipek2009} when organizing the findings. The notion of infrastructure frames ICT systems that have various functions as the foundation of users' activities; infrastructure is invisible to users until it breaks down \cite{Star1996}. When it breaks down, users' work to address the breakdowns is called "infrastructuring work \cite{Pipek2009}." The first theme illustrates how blind users employed TikTok for various purposes, encompassing everyday entertainment, professional development, and community engagement. The second theme highlights the challenges and solutions encountered by blind users while utilizing TikTok. Despite viewing TikTok as an infrastructure, BlindTokers faced breakdowns due to inaccessible designs. Consequently, they undertook infrastructuring work to address the accessibility breakdowns. Further elaboration on these themes is presented in the findings section. Additionally, we observed that participants' perceptions of TikTok were closely tied to the concept of independence, a crucial aspect in the field of accessible computing \cite{10.1145/3449223}. We then discussed the findings with previous literature on independence and infrastructuring work. We first unpacked the nuanced nature of independence with a multilevel perspective; we then critically examined the infrastructuring work carried out by participants for independence. We also provided design implications for a more accessible TikTok.

This study has several contributions to the literature: 1) it documents blind users' various activities on TikTok, adding empirical examples to the literature on how blind people use short-video platforms; 2) it unpacks blind users' contextualized understanding of independence while using TikTok; 3) it also critically examines the TikTok infrastructure, including its accessibility breakdowns as well as its potential ableist culture.

\section{Related Work}

In this section, we review three lines of literature that are relevant to our study. First, we review the literature on infrastructure and infrastructuring work. In this subsection, we frame large-scale ICT systems for a large variety of purposes, such as TikTok, as "infrastructure" to denote the systems' complexity and pervasiveness; we then focus on the disruptions users experience when using the infrastructure and their work to mitigate the disruptions. Second, we narrow down the focus of the review to research on independence in accessible computing to demonstrate how HCI researchers study the inner desires of users with disabilities. Third, we survey the recent literature specifically on blind users and social media to understand how people with visual impairments use social media. We highlight the accessibility issues that specifically affect blind users' experiences.


\subsection{Infrastructure, Infrastructure Disruptions, and Infrastructuring Work}

First, we draw on the existing literature on infrastructure and infrastructuring to provide a framework for understanding the interactions of blind users with TikTok. In the HCI community, infrastructure is conceptualized as a social-technical system of large-scale that supports users' daily routines \cite{Star1996, Neumann1996, Star1999, Bowker1999, Edwards2003}. Infrastructure can serve as the foundation for various citizen activities, such as public libraries \cite{Kozubaev2021}, or ICT platforms facilitating a wide range of online activities like WeChat \cite{Zhou2020}. Infrastructure predominantly operates behind the scenes, as noted by Star and Ruhleder \cite{Star1996}, who emphasized that when infrastructure functions smoothly, it "\textit{sinks into an invisible background. It is something that is just there, ready-to-hand, completely transparent} (p.112) \cite{Star1996}." Consequently, the invisibility of infrastructure often leads users to take it for granted \cite{10.1145/2998181.2998344}, and its existence may go unnoticed until a breakdown occurs \cite{Kaziunas2019, Bhat2021}. When infrastructure experiences a breakdown, users' routines are disrupted, and they must independently address the issue themselves. This remedial effort to resolve infrastructure malfunctions is termed "\textit{infrastructuring work} \cite{Pipek2009}."

Infrastructure researchers have extensively studied the disruptions that occur in users' everyday lives, particularly when these disruptions are ongoing \cite{10.1145/3290605.3300688, semaan2011, Mark2009, 10.1145/3491102.3502126, 10.1145/3491102.3501980}. Some of these ongoing disruptions stem from prolonged and violent conflicts that damage or destroy infrastructures, such as wars \cite{10.1145/3491102.3502126}. Additionally, other ongoing disruptions arise from the marginalization and exclusion of specific populations, resulting in their lack of access to infrastructure and constant disruptions in their lives. For instance, Semaan et al. \cite{10.1145/2858036.2858109} conducted a study highlighting the lack of infrastructural support for retired US veterans, which led to significant challenges for their transition back to civilian society. Recently, researchers have been paying growing attention to algorithmic exclusion, the infrastructural exclusion in large-scale algorithmic system contexts. Algorithmic exclusion manifests as "\textit{the ways in which algorithms construct and reconstruct exclusionary structures within a bounded sociotechnical system, or more broadly across societal structures.} (P.24)\cite{10.1145/3432951}" The exclusion foregrounded algorithmic system's (e.g. TikTok) inherent discrimination and marginalization of some populations, such as LGBTQ+ people, through various algorithmic functions.

The goal of infrastructuring work is to address infrastructure failures, although the specific outcomes of such work are context-dependent, varying based on the victims' situations \cite{10.1145/3461778.3462075, 10.1145/2493432.2493497, 10.1145/3359175, 10.1145/3173574.3174055, 10.1145/2493432.2493497}. A group of researchers has taken an interest in the infrastructuring work carried out for the purpose of well-being when facing infrastructure disruptions \cite{10.1145/3359175, 10.1145/3461778.3462075}. For instance, Semaan \cite{10.1145/3359175} highlighted the significance of building "\textit{resilience}" in response to prolonged infrastructural disruptions. Resilience, as defined in the study, refers to "\textit{the practices people develop to bounce back from, manage, and overcome disruption in their daily lives} (p.2) \cite{10.1145/3359175}." The study introduced the concept of "routine infrastructuring" as an approach to building resilience. This involves engaging in constant and repetitive technology-supported practices that assist victims in reconfiguring their lifestyles, thus enabling them to better cope with and gain confidence in facing disruptions. Specifically on TikTok, Simpson \cite{simpson2023hey} reported a type of infrastructuring work that involved content creation. It is called "\textit{creative labor}," which refers to "\textit{the time, effort, and creativity needed to produce an artistic output} (p.7) \cite{simpson2023hey}." When it comes to infrastructuring work specifically for accessibility, Boujarwah et al. \cite{10.1145/2049536.2049542} emphasized: "\textit{(people with disabilities)...in order to live independently,... it was apparent that having computer skills and other relevant simple technology competency is important} (p.21) \cite{10.1145/2049536.2049542}".

In this paper, we add to the literature on the infrastructuring work of people with disabilities. Specifically, we take users with visual impairments as an example and investigate how they used TikTok as infrastructure for various online activities. We also explore how blind users conducted infrastructuring work to address the accessibility breakdowns of TikTok.

\subsection{Independence, Accessibility, and HCI}

We also use independence as the lens to explore the visually impaired users' motivations while using TikTok. Independence is an important concept in HCI \cite{10.1145/3449223,10.1145/3290607.3312829,10.1145/3290607.3312840}, Accessibility \cite{REINDAL1999,Storer2021}, Ageing \cite{Hillcoat-Nalletamby2014,Breheny2009}, and Rehabilitation \cite{Dixon2007}. It has many dimensions. According to previous HCI studies on accessibility, \cite{10.1145/3449223,10.1145/3274354}, independence can be interpreted as a combination of \textit{self-reliance, self-efficacy, control, autonomy, reciprocity} and \textit{interdependence}. Self-reliance, self-efficacy, control, and autonomy are related to freely choosing and finishing their own tasks. Self-reliance refers to the sense of "\textit{'caring for yourself,' without getting help from others} (p.2) \cite{10.1145/3449223}"; self-efficacy is one's "\textit{belief about his or her capabilities to perform an activity that has influence over events affecting his or her life} (p.231). \cite{Dixon2007}"; control means "\textit{the capacity to perform behaviors needed to influence outcomes in their environment.} (p.633) \cite{Wehmeyer1996}"; autonomy is the "\textit{self-determination in choosing what to do, how and when to do it} (p.422) \cite{Hillcoat-Nalletamby2014}." In addition, reciprocity and interdependence help people be independent while developing or maintaining social relationships with others. Reciprocity is "\textit{an equivalence of giving and receiving... (and it gives) people a clear rejection of the position of dependency through the maintenance of equal relationships.} (p.1308) \cite{Breheny2009}"; interdependence emphasizes "\textit{multiple forms of assistance happening simultaneously} (p.164) \cite{Bennett2018}," foregrounds "\textit{the often-underwritten contributions of people with disabilities} (p.164) \cite{Bennett2018}," and overcomes "\textit{hierarchies that prefer ability} (p.164) \cite{Bennett2018}."

In the field of HCI, there has been extensive research on the impact of technology on the independence of various populations with disabilities, including individuals with visual impairments \cite{10.1145/3449223, 10.1145/3274354, 10.1145/1978942.1979424, 10.1145/3290605.3300246} and individuals with hearing disabilities \cite{10.1145/1978942.1979424}. While many studies highlight how technology enhances independence by providing assistance to users \cite{10.1145/3290605.3300246}, some scholars have also identified instances where technology can hinder independence \cite{10.1145/3449178, 10.1145/3479607, Caldeira2022}. When technology provides excessive assistance, it can inadvertently undermine users' independence by making them feel infantilized or stereotyped \cite{10.1145/3449178}. Therefore, researchers emphasize that the role of technology in improving independence is contextual, necessitating the users' active efforts to identify, navigate, and manage how, where, and what aspects of their lives can become more independent with the aid of technology \cite{guldenpfennig2019autonomy}. As noted by Lee et al. \cite{10.1145/3449223}, users perceive technology-mediated independence as "\textit{not a static or slow-changing personality trait, but as an experience that they attain to different extents through their choices, skills, and tools} (p.6) \cite{10.1145/3449223}." Furthermore, researchers underscore the importance of collaboration with others in achieving independence \cite{10.1145/3449143, 10.1145/3274295}. In a study examining the collaboration among patients, caregivers, and clinicians, Buyuktur et al. \cite{10.1145/3274295} argued that independence "\textit{must be co-constructed by the choices and activities of the care network, including the person with a disability, caregivers, and clinicians} (p.1) \cite{10.1145/3274295}."

However, most studies on independence in the current HCI literature on accessibility primarily revolve around task-oriented activities like shopping and traveling, with limited exploration of how users navigate social media to enhance their independence. An exception to this is Hong's studies on the use of social network sites (SNS) by young adults with autism \cite{10.1145/2470654.2466439, 10.1145/2145204.2145300}. In the first study, Hong et al. \cite{10.1145/2145204.2145300} highlighted that due to the peripheral role of social media in the lives of participants, expressing the need for support on social media was more informal, thus fostering a sense of reduced dependence. The second study \cite{10.1145/2470654.2466439} shed light on the dilemma faced by young adults with autism, as their social networks were typically shaped by their parents, resulting in limited independence in socializing. To address this, the research team proposed an online service that enabled users to create their own networks based on trusted real-life friends. Additionally, users had the option to expand their network by allowing their trusted friends to add more individuals they deemed trustworthy. These networks facilitated increased independence for young adults with autism, granting them the freedom to communicate within a larger network of trusted members. However, it is important to note that these studies primarily focused on social networks comprised of trusted individuals and did not explore social relationships with strangers, nor did they elucidate the specific contextual conditions in which users must navigate social-technical resources online to establish or maintain their independence.


In this study, we fill the gap by presenting a study on how BlindTokers constructed independence and what role TikTok, a famous social media platform, played in the process of constructing independence. Specifically, we are interested in unpacking BlindTokers' nuanced perceptions of independence (i.e. how they understand independence in various contexts while interacting with TikTok).

\subsection{Blind Users, ICTs, and Social Media}

Finally, our study builds upon prior research examining the experiences of users with visual impairments on social media platforms. HCI and Accessible Computing have conducted numerous studies on the use of information and communication technologies (ICTs) by individuals with visual impairments \cite{Brady2013, 10.1145/3132525.3134801, 10.1145/3308561.3353792, 10.1145/3025453.3025814, 10.1145/3441852.3476550}. Due to accessibility issues, individuals who are blind often encounter difficulties in accessing visual information through digital technologies \cite{Abdolrahmani2016a, Jaramillo-Alcazar2017, 10.1145/3170427.3170633}. As a result, HCI research has primarily focused on developing more inclusive designs that facilitate improved screen reading experiences for blind individuals \cite{10.1145/3234695.3241025, 10.1145/3313831.3376267, Jandrey2021, 10.1145/3411764.3445233}. Popular approaches include the use of screen readers to audibly present text or images on screens \cite{10.1145/3411764.3445233} and computer vision technology to detect and interpret real-life objects \cite{Wu2014}. Nevertheless, challenges persist. Regarding text information, a notable issue is that screen readers may not be able to identify unlabeled buttons \cite{Ross2018}. In the case of pictures, aside from the prevalent lack of alternative text for most images \cite{10.1145/3368426}, researchers have also observed that blind individuals' focus on images is situated and challenging to describe \cite{Abdolrahmani2016, 10.1145/3313831.3376404, Jung2022}. In addition to addressing these challenges in accessing visual information, researchers have also explored the intrinsic desires of blind users when utilizing ICTs \cite{Li2022, 10.1145/3290605.3300602, 10.1145/3411764.3445321, Stangl2022}. They have investigated how these desires, including being ordinary ("\textit{to appear as ordinary members of the society, unnoticeable and not be treated differently} (p.2)") \cite{10.1145/3290605.3300602}, dignified \cite{10.1145/3411764.3445321}, beautiful \cite{Li2022}, competent \cite{10.1145/3313831.3376767}, and independent \cite{10.1145/3449223}, are mediated through the use of ICT.

The literature has documented numerous studies on the use of social media platforms by individuals who are blind, such as Twitter \cite{10.1145/3308561.3354629,10.1145/3313831.3376728} and Facebook \cite{Voykinska2016}, where the content primarily consists of text and images. Recently, there has been an increasing focus on video-oriented platforms used by blind users \cite{Lee2021a, Seo2021, Seo2018, Seo2021a}, where they create or consume video clips and participate in live-streaming on platforms like YouTube or Twitch. These platforms feature experiences of consuming video content, which might "\textit{impose additional burdens and challenges on people with visual impairments.} (p.2) \cite{10.1145/3476038}". Jun et al. \cite{10.1145/3476038} examined the interactions of people with visual impairments with live-streaming services. They found that participants encountered challenges in accessing information on video-oriented platforms. For example, blind streamers reported difficulties in reading the floating comments left by viewers on the screen. To overcome these challenges, they relied on various socio-technical resources, such as external screen readers or assistance from other sighted people to read the chat messages. Among the studies, a recently published paper on blind users' experiences with Douyin (the Chinese version of TikTok), by Rong et al. \cite{Rong2022}, is particularly relevant to this research. They investigated users' perceptions and interactions with Douyin's content curation algorithms. The study highlighted the challenges faced by users due to the platform's inaccessible designs when streaming or creating content on Douyin. It specifically revealed the tensions between blind and sighted users on the platform and the strategies employed by blind users to sidestep these conflicts.

In this study, we also join the inquiry of blind users' experiences on short video platforms. It should be noted that our paper is different from Rong's study \cite{Rong2022} in two primary ways: 1) The first difference is about the research focus. The previous pivots around the interactions between blind users and the platform. However, our study is centered around BlindTokers' various interactions, not just the user-platform interaction. 2) The second one is about the context of the study. While Rong's study focuses on the Chinese context, our paper is about people in the US, Canada, and the UK. The cultural backgrounds are different. The difference also provides rich implications for the literature on HCI and Accessible Computing, which we will cover in the future work section.

\section{Methodology}
In this section, we present the TikTok designs that are relevant to the experiences of blind users. We also describe the data collection and analysis methods employed in this study. The research team obtained IRB approval from their institution prior to data collection. To answer the research questions, we conducted a qualitative study. This study involved semi-structured interviews and thematic analysis as the research methods. 

\subsection{Study Site: TikTok Design and Blind Users}

\begin{figure}[htp]
    \centering
    \includegraphics[width=15cm]{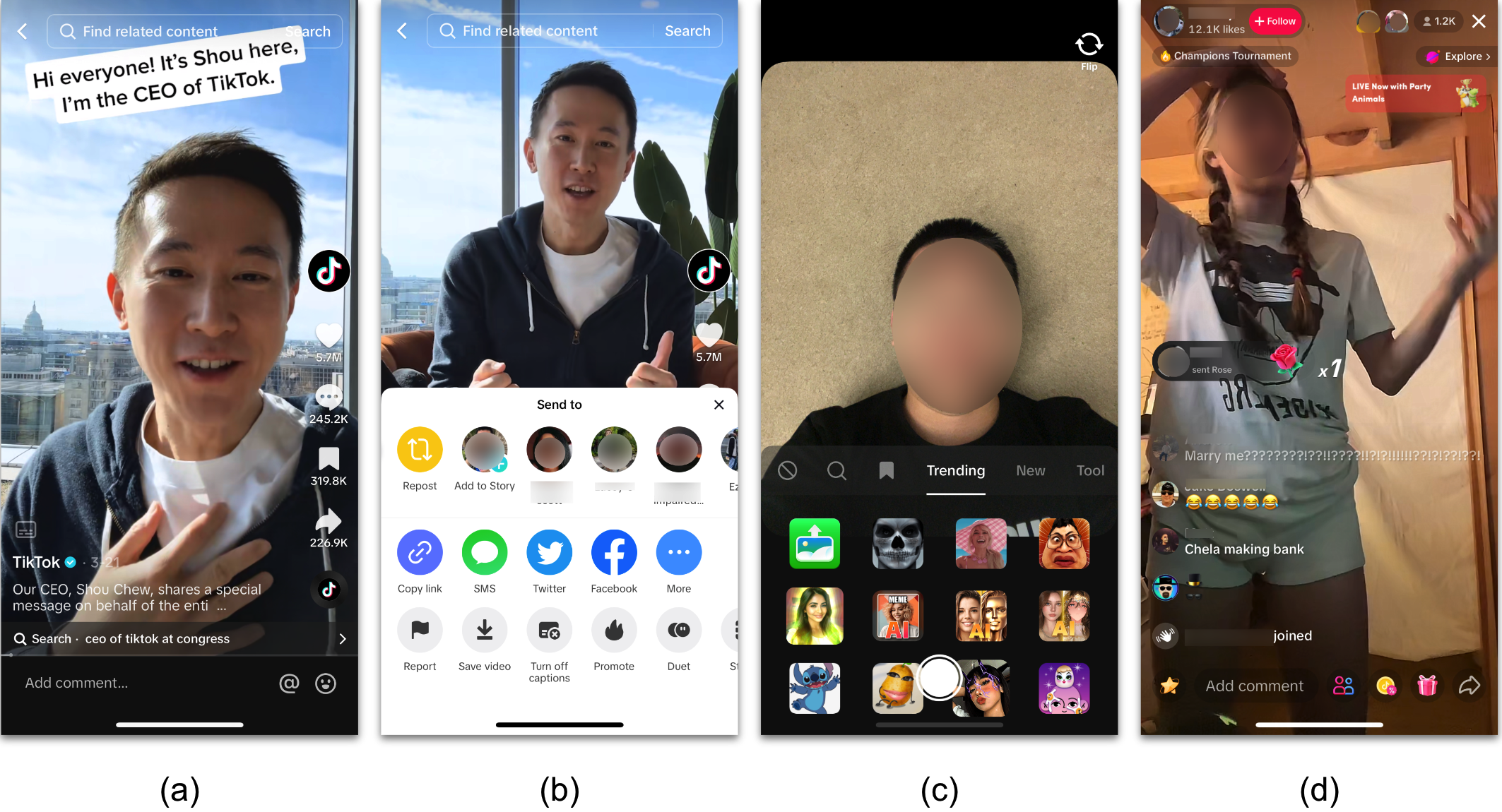}
    \caption{TikTok's Interfaces of Video Presentation (a), Sharing Functions (b), Video Filters (c), and Live-streamings (d) on iPhone.}
    \label{fig:1}
\end{figure}

TikTok is a large platform that offers a wide range of services. We demonstrate the functions and designs that are relevant to the experiences of blind users. The main interface of TikTok features a video that occupies most of the screen. The buttons, descriptions, and tabs are positioned on the left, bottom, and top of the video (Figure \ref{fig:1}, a). Users can share content through links, SMS, or messages (Figure \ref{fig:1}, b). On the video creation page, users can utilize their phone camera to record videos and apply visual and/or audio effects (Figure \ref{fig:1}, c). Users can also host live-streaming. The interface displays real-time comments and gifts from viewers, and the lower part of the screen automatically scrolls to show the most recent comments or gifts (Figure \ref{fig:1}, d). To enhance the experience of consuming videos, creators can add captions to their videos (Figure \ref{fig:1}, a).




Therefore, TikTok is a highly visual-oriented platform. Many user experiences on the platform revolve around recognizing and engaging with text, color, and images. To maximize the display of image or video content on the screen, TikTok opts for small, low-contrast, and relatively transparent text fonts, icons, and buttons. This design choice aims to prevent them from overlapping with the visual content. However, the impact of TikTok's design on blind users' experiences was significant, as reported by the participants. A comprehensive description of blind users' experiences on TikTok will be provided in the findings section.




\subsection{Recruiting Participants}

All participants were recruited from TikTok through direct messages and comments. It is important to note that TikTok has strict restrictions on direct messages between strangers. Users can only send direct messages to each other if they mutually follow each other. Therefore, the participant recruitment process consisted of four stages. 1) The research team registered official accounts on TikTok, providing their real identity information and clearly stating the purpose of the project. They searched for keywords like "blind," "visual impairment," "low vision," etc., to identify TikTokers who openly identified themselves as visually impaired people in their profiles or videos. A list of such users was created. 2) The team followed all visually impaired users on the list and actively engaged with them in a sincere, polite, and honest manner. They posted videos introducing themselves, asked questions about how users used and perceived TikTok, attended live streams, shared life experiences, and left comments and "likes" to support the participants' TikTok activities. 3) Through a series of interactions, the team established trust and interest with some users, who then followed the team's TikTok accounts back. This allowed for more in-depth conversations through private messages and successful invitations to participate in interviews. 4)At the end of the interviews, participants were asked to recommend other TikTokers with visual impairments. Through this snowball sampling process, more participants were recruited.


To ensure that the participants belonged to our intended population, we verified during the interviews that they identified as visually impaired and were active users of TikTok. Additionally, we exclusively interviewed individuals who were at least 18 years old and proficient in English.

Finally, we recruited 30 informants, including 2 non-binary people, 13 females, and 15 males. All gender information was self-reported by participants without any pre-defined options. The ages of the interviewees ranged from 19 to 59 years (avg. 34.9). Most of the participants (27 out of 30) came from the United States, one came from the United Kingdom, one from Canada, and one did not provide residency information. Among the participants, 12 had high school education, 2 had general education degrees, 7 had associate degrees, 6 had bachelor's degrees, 2 had master's degrees, and 1 had a Ph.D. degree. In terms of occupation, 6 of them were unemployed, 6 were college students, and 18 of them had a large variety of professional backgrounds (e.g., musician, therapist, financial consultant). In terms of visual impairments, 22 were legally blind, 7 had low vision, and 1 was completely blind; 15 of the blind participants were diagnosed more than 10 years ago, 10 more than five years ago, and 5 of them were diagnosed in recent five years. Speaking of using TikTok, on average, 20 users spent more than an hour on TikTok per day, 7 spent 30-60 minutes per day, and 3 spent less than 30 minutes per day. To avoid potential privacy and anonymity issues, we only report information about age (range), visual impairments, time since onset, daily usage of TikTok, and occupation in Table ~\ref{table:demo}.

\begin{table}[]
\caption{Demographics of Participants}
\begin{tabular}{cccccl}
\hline
\textbf{\#} & \multicolumn{1}{c}{\textbf{Age (range)}}  & \multicolumn{1}{c}{\textbf{Impairment}} & \multicolumn{1}{c}{\textbf{Since Onset}} & \multicolumn{1}{c}{\textbf{Daily Usage}}& \multicolumn{1}{l}{\textbf{Occupation}} \\ 
\hline
P01 & 18-30 & Low Vision    & 27 yrs & >60 mins  & Student              \\
P02 & 18-30 & Low Vision    & 24 yrs & >60 mins & Student              \\
P03 & 18-30 & Legally Blind & 1 yr   & >60 mins  & Student              \\
P04 & 18-30 & Legally Blind & 25 yrs & >60 mins  & Unemployed           \\
P05 & $\ge$51 & Legally Blind & 5 yrs  & >60 mins & Software Developer   \\
P06 & 31-50 & Legally Blind & 39 yrs & >60 mins & Tech Advisor         \\
P07 & 31-50 & Legally Blind & 32 yrs & >60 mins & Unemployed           \\
P08 & 31-50 & Legally Blind & 29 yrs & >60 mins & Music Composer       \\
P09 & 31-50 & Legally Blind & 15 yrs & 30-60 mins  & XR Designer          \\
P10 & 31-50 & Low Vision    & 32 yrs & >60 mins & Musician             \\
P11 & $\ge$51 & Legally Blind & 4 yrs  & 30-60 mins  & Gallary Owner        \\
P12 & 18-30 & Legally Blind & 21 yrs & 30-60 mins  & Unemployed           \\
P13 & $\ge$51 & Totally Blind & 59 yrs & >60 mins & Finacial Consultant  \\
P14 & 31-50 & Low Vision    & 4 yrs  & 30-60 mins & Car Dealer           \\
P15 & $\ge$51 & Legally Blind & 53 yrs & >60 minss & Lawyer               \\
P16 & 31-50 & Legally Blind & 40 yrs & >60 mins & Church Minister      \\
P17 & 31-50 & Legally Blind & 31 yrs & >60 mins & Music Composer       \\
P18 & 31-50 & Legally Blind & 21 yrs & 30-60 mins & Fitness Instructor   \\
P19 & 31-50 & Low Vision    & 2 yrs  & >60 mins & Sales Manager        \\
P20 & 18-30 & Low Vision    & 15 yrs & >60 mins & Unemployed           \\
P21 & 31-50 & Legally Blind & 12 yrs & >60 mins & Unemployed           \\
P22 & 18-30 & Legally Blind & 22 yrs & >60 mins & Student              \\
P23 & 18-30 & Legally Blind & 28 yrs & >60 mins & Therapist            \\
P24 & 18-30 & Legally Blind & 29 yrs & <30 mins & Motivational Speaker \\
P25 & 31-50 & Legally Blind & 9 yrs  & >60 mins & Tech Expert          \\
P26 & 18-30 & Legally Blind & 15 yrs & <30 mins & Digital Marketing    \\
P27 & 18-30 & Low Vision    & 30 yrs & <30 mins & Unemployed           \\
P28 & 31-50 & Legally blind & 19 yrs & 30-60 mins & Student              \\
P29 & $\ge$51 & Legally Blind & 20 yrs & >60 mins & Tech Advisor         \\
P30 & 18-30 & Legally Blind & 19 yrs & 30-60 mins & Student               \\
\hline       
\end{tabular}
\label{table:demo}
\end{table}

\subsection{Designing Interview Protocol}
To understand the experiences of TikTokers with visual impairments, we developed a semi-structured interview protocol. In the protocol, we focus on three primary interactions (to foreground the accessibility issues of TikTok, we also ask follow-up questions like "Is this accessible?" and "How do you address the accessibility issues?" after each question): 

\begin{enumerate}
\item \textbf{The overall perceptions of TikTok.} Example Questions: "\textit{What are the main purposes of using TikTok?}" "\textit{What are the differences between TikTok and other popular social media platforms (e.g. Instagram, YouTube, Twitter, and/or Facebook.)?}" 

\item \textbf{The experiences of video creation and consumption on TikTok.} Example Questions: "\textit{How to make videos on TikTok?}" "\textit{How to host live-streaming?}" "\textit{What videos/live-streaming do you like most on TikTok}" \textit{"How to make duet/stitch videos?"} \textit{"What are the experiences of consuming TikTok videos?"}.

\item \textbf{The experiences of non-video interactions on TikTok.} Example Questions: "\textit{How to leave comments on TikTok}" "\textit{How to share videos?}" "\textit{How to send private messages?}".

\end{enumerate}



When interviewing participants, we carefully listened to participants' descriptions of such interactions and experiences. We especially paid attention to the challenges that participants encountered due to inaccessible designs; we then went further and explored what role TikTok played in the challenges, like "did it create such challenges or helped participants overcome the challenges"; we also showed particular interest in how participants came up with the solutions for the challenges. Due to the diversity of participants' experiences, we cannot include all the scenarios and challenges in the interview protocol. Therefore, in addition to the pre-designed questions, we used prompts and follow-up questions when we found interesting experiences. 

Finally, we interviewed 30 participants, and the interviews ranged from 25 to 60 minutes (avg.=41.5 mins). Taking into account the diversity of blind users on TikTok and the situation of the COVID-19 pandemic, all interviews were conducted remotely. Most of the interviews were conducted through Zoom; few participants attended the interviews by phone calls, due to the accessibility issues of Zoom. Before each interview, we clearly informed the participants about our research team and the purpose of the project. We started interviewing them after getting their verbal consent. To better capture the details of their experiences, we audio-recorded the interviews and then transcribed them into text documents. For readability reasons, we modified the quotes when we presented them in the Findings section, such as adding context information marked with "()," and removing meaningless utterance words like "umm."   


\subsection{Data Analysis}


We used thematic analysis \cite{Braun2006b} to analyze the data. Thematic analysis is a type of method that comprehends data by breaking it down into meaningful segments, looking for patterns across segments, and organizing the segments into themes based on the patterns. Particularly, there are six steps of thematic analysis \cite{Braun2006b}: 1) familiarize with data, 2) generate initial codes, 3) search for themes, 4) review themes, 5) define and name themes, and 6) produce the report. Both authors participated in the data analysis; we met weekly to discuss the coding results and resolve disagreements. We followed the procedure to ensure the validity and reliability of data analysis: 

\begin{enumerate}
  \item We got ourselves immersed in the data. We read through all the interview transcripts to obtain an overall understanding of the participants' experiences with TikTok. We took notes that helped us identify potentially meaningful interactions between participants and TikTok.
  \item After getting familiar with the data, we got an overall impression that participants used TikTok for various purposes, but the experiences were hindered sometimes due to accessibility issues. Especially, they reported their work in terms of addressing accessibility issues. Therefore, we looked for all the quotes that described such interactions and coded them. We got 121 meaningful codes (examples of codes are shown in \ref{tab:code}).
  \item We documented all the codes and tried to group them into exhaustive and exclusive themes. To be consistent with the research questions, we chose to group the codes into two primary groups: one for participants' purposes of using TikTok and one for challenges and solutions when using TikTok.
  \item We carefully reviewed the themes. In this process, we noticed that participants' purposes for using TikTok were various, but generally could be categorized as "entertainment" "professionalization" and "community." The solutions to accessibility challenges also can be classified as "technical work" and "social interactions."
  \item When defining the themes, we noticed BlindTokers' use of TikTok was highly relevant to infrastructure \cite{Star1996}, a concept in HCI that describes ICT-based socio-technical foundations for user experiences. Besides, blind users' work to address the accessibility problems was also related to infrastructuring work \cite{Pipek2009}, a concept that investigates users' work to fix the breakdowns of infrastructure. Therefore, we decided to use the terminology based on these concepts to name the themes. We also made sure the refined themes were consistent with the research questions. We finally got two overarching themes: "TikTok As Infrastructure For Blind Users’ Multiple Purposes" and "TikTok’s Accessibility Breakdowns and BlindTokers’ Infrastructuring Work."
  \item We reported all the themes and subthemes with proper examples and interpretations in the findings section.
\end{enumerate}


\begin{table}[htbp]
    \centering
    \caption{Example of Code in "Community Engagement"}
    \renewcommand{\arraystretch}{2}
    \begin{tabularx}{\textwidth}{|p{\dimexpr 0.2\textwidth-2\tabcolsep} | p{\dimexpr 0.3\textwidth-2\tabcolsep}|p{\dimexpr 0.5\textwidth-2\tabcolsep}|}
        \hline
        \multicolumn{1}{|c|}{\textbf{Code Name}}
         & \multicolumn{1}{c|}{\textbf{\textbf{Description}}} & \multicolumn{1}{c|}{\textbf{\textbf{Example}}} \\
        \hline
        \textbf{Feeling related to other blind users} & BlindTokers gained the sense of relatedness when they consume other blind users' videos & "\textit{...it gives you a real a sense of other people going through the same kind of issues ...and living their day to day life.}[P14]" \\

        \textbf{Supporting blind creators' content} & BlindTokers supported other blink users' content with TikTok functions &  "\textit{I will be showing support by liking and replying to them... you can reply to that comment with the video so that people can see the comment.} [P08]" \\
        
        \textbf{Defending blind friends} & BlindTokers defended their bind friends from harassment on TikTok & "\textit{...she will get the complete jerk saying 'you're faking...you are not blind.'...I just comment underneath...in a very, not nice way. If you're gonna be a jerk to my friend, I'm gonna tell you off.} [P20]"  \\
        \hline
    \end{tabularx}
    \label{tab:code}
\end{table}

\subsection{Researcher Positionality and Ethical Considerations}

Finally, we state the positionality of the researchers of this study. The research team involved in this study consists of multiple researchers, all of whom are sighted. The first author, an indigenous Asian male, played a primary role in conducting this project. Prior to this study, he participated in several initiatives aimed at enhancing the accessibility of ICTs for blind individuals. Additionally, he has been actively involved as a volunteer within the local community of blind residents. He got familiar with blind people's lives through various activities like facilitating experiments in research projects or providing everyday assistance. The last author, a white male, is the advisor of the first author; he has led multiple projects focusing on addressing various ICT accessibility issues for blind people.

We fully acknowledge the potential bias that may arise from the absence of researchers with direct experiences of visual impairments within the team. In order to mitigate this bias, we strictly followed protocols during the data collection process. Instead of using printed flyers, we used TikTok videos to promote the project and recruit participants. In these videos, the researchers presented their real identities, including their names, affiliations, email addresses, and facial information, to establish transparency and credibility. We reached out to participants based on their self-disclosures on TikTok, and prior to the interviews, We politely confirmed the participants' visual impairments without imposing any preconceived criteria. We gathered demographic information through verbal interactions rather than requesting participants to fill in forms. Furthermore, we gave careful consideration to the language used in communication with participants, such as rephrasing "see you then" to "meet you then" and "how do you watch videos" to "how do you consume videos". In light of privacy concerns on video-sharing platforms \cite{lookingbill2022examining}, we only reported brief demographic information of participants in the paper. When writing the paper, we also carefully phrased our language to avoid exclusion or othering. Despite the efforts, we acknowledge the limits of this study due to the background of the research team.
\section{Findings}

We present our findings to answer the two research questions. The first subsection describes how blind users utilize TikTok for various purposes. Based on the accounts of the participants, TikTok plays a significant role in their lives, encompassing various aspects of their daily activities, including everyday entertainment, professional development, and community engagement. The extensive coverage makes TikTok an essential infrastructure in the daily lives of BlindTokers. The second subsection showcases the challenges and solutions encountered when using TikTok. Although blind users consider TikTok as part of their infrastructure, it breaks down due to inaccessible designs. These breakdowns hinder blind users' content consumption experiences and limit their ability to create creative content. BlindTokers have to devise solutions, including utilizing technical skills and leveraging social relationships, to cope with accessibility breakdowns.

\subsection{TikTok As Infrastructure For Blind Users' Multiple Purposes}


This subsection addresses the first research question by examining the different ways participants utilized TikTok. The first aspect is everyday entertainment, where participants engaged with TikTok as a form of leisure and distraction. The second aspect involves using TikTok as a tool for professional development, including enhancing skills and promoting businesses. The third aspect focuses on community engagement, where participants established communities specifically for blind users ("BlindTok") and formed alliances with other minority groups on TikTok. In these instances, TikTok functioned as a socio-technical infrastructure, facilitating and supporting the activities of blind users through its diverse features and designs. However, it is important to acknowledge that TikTok was not without accessibility problems, which will be discussed in the following subsection.

\subsubsection{\textbf{TikTok for Everyday Entertainment}}

All participants (N=30) reported using TikTok for entertainment during their leisure time. Entertainment served as a means for them to consume and create content for enjoyment. Overall, participants found the consumption of TikTok content to be convenient. Compared to other social media platforms that focused on images or text, TikTok, with its video-centric format, provided more accessible information. Specifically, a significant number of videos on TikTok included audio narratives, which audibly explained the content to blind users ([P14]). This feature made TikTok's content more accessible compared to that on Instagram or Facebook. Additionally, TikTok's browsing functions, such as endless scrolling and double-tapping to "like," were intuitive and easy to grasp. Participant P19 shared how he utilized TikTok to access more content that was friendly to blind users:

\begin{quote}\textit{I was trying to curate my "for you page" into content that I enjoy, orally based. If I come across videos mostly text, I immediately skip. I follow blind creators, I'll throw a comment and share with my wife, to promote their content, so you can see more. [P19]}\end{quote}

P19 showed how to leverage the algorithmic features to curate personalized content on TikTok by conducting various interactions, echoing previous work \cite{Rong2022}. In many cases, such accessibility and convenience made TikTok a relatively reliable application for browsing blind-friendly content in leisure time.  


In addition to content consumption, participants also shared their experiences of creating content on TikTok. Nearly all participants reported using TikTok to record and share their daily lives, particularly from the perspective of visually impaired individuals. Some participants, like P15, utilized TikTok videos to illustrate the unique aspects of their visual impairments to others:

\begin{quote}\textit{Because of my vision, I tend to focus into the middle. When I use that portrait filter...it helps others (understand) it is just the way I see the world. [P15]}\end{quote}

According to the participants, there were many aspects of blind people's lives that sighted individuals were not aware of, possibly due to a lack of education. As a result, the participants took it upon themselves to educate the public about the experiences of being blind. One common educational topic was the understanding that "\textit{blindness is such a wide spectrum} [P07]". Being blind does not necessarily mean a complete loss of vision, as visual impairments can vary. P15 showcased his visual impairment using TikTok's filters. Additionally, he praised TikTok's content creation in comparison to YouTube:


\begin{quote}\textit{I used to try to make YouTube video, but it was such a production, I don't have time for it. But for TikTok...like I often "ding" and something comes into my head. I just do it on the way, posted the draft and finish when I get home. [P15]}\end{quote}


TikTok has gained popularity for its vlogging-style videos. In these vlogs, users typically hold their cell phones, record selfie videos for several minutes, and engage in casual conversation. Based on P15's comments, this "talking head" style actually made content creation easier for blind users. When comparing video styles between TikTok and other platforms, participants further expressed that the ease of creating content on TikTok contributed to feeling less dependent. According to P07, \textit{"I suppose TikTok is much easier. I just don't make fancy videos so much so I don’t need to ask for help."} 

Besides the characteristics of visual impairments, some participants showed a large variety of their everyday activities as people living with visual impairments in their videos. P06 told us: 

\begin{quote}\textit{...because most of us, we don't really shout out about it, kind of keep it very much to ourselves. This is my way of not hiding away from it. A lot of videos show me about actually doing things, having a job, working, the fact to have children, managing successful relationships, showing them being independent...[P06]}\end{quote}

P06's narrative highlighted how TikTok provided a platform where she could confidently share her experiences of living with visual impairments. The narrative also illustrated that participants sought not only to live independently while using TikTok but also found motivation to showcase their autonomous lives on the platform. However, some participants, such as P16, expressed apprehension about the potential of over-presentation on TikTok. This concern was articulated by citing a TikToker with visual impairments' introductory line frequently used in their videos:


\begin{quote}\textit{...there's a blind college boy, like (exaggerated voices) "YO, I’M BLIND, LET’S SEE HOW I COOK!" and that's how he starts all his videos...But for a lot of blind people, it actually could be a little problem, their parents could come to them and say, "wow, (they) could do it, why can't you?"}\end{quote}


P16 voiced apprehension that this kind of portrayal might foster an environment encouraging individuals with disabilities to exhibit their capabilities on social media. Rather than inspiring, this environment could potentially cause discomfort for those with disabilities.


\subsubsection{\textbf{TikTok for Professional Development}}


While many participants viewed TikTok as a platform for everyday entertainment, a significant number of them (N=15) employed it for professional growth. Specifically, these participants demonstrated a greater interest in using TikTok to refine their professional skills or promote their businesses.

The first type of professional development was proficiency training. Many interviewees in our project were employed, offering their expertise and skills to the benefit of society. They harbored concerns about their professional trajectory and desired to enhance their proficiency. According to several participants (N=7), they were interested in harnessing TikTok as a tool for practicing skills. Some turned to TikTok for learning. P06 recounted how she used the platform to assist her daughter, who was also blind, in learning Braille:


\begin{quote}\textit{My daughter, she's fiercely independent. She explores and she's a tough little cookie. And she's supposed to be learning Braille. So I bought simple things people from TikTok recommended, like a dice, but it's got Braille on each side. You can pop it to do the alphabet and start learning Braille. She quite enjoys that. [P06]}\end{quote}


While reading Braille may not be directly tied to specific professional skills, it holds considerable significance in the lives of blind people. Therefore, learning to read Braille was important for P06's daughter. The manner in which P06 obtained the dice also illustrated how TikTok could serve as a beneficial platform for blind people to exchange advice, particularly on experiencing life with visual impairments. Apart from learning-related skills, participants also reported developing leadership-related skills. P11 stated:

\begin{quote}\textit{And some of my videos were just me trying to get over my intimidation of speaking in front of people, because I was elected the president of our local [blind organization] [P11] }\end{quote}


As a leader, delivering a speech in front of an audience necessitates diverse skills. P11 shared with us that while she initially had a reserved demeanor, she had to enhance her public speaking skills as she was stepping into a leadership role within an organization for the blind. She utilized videos on TikTok as a method to foster her confidence and improve her speaking abilities. Moreover, some participants incorporated TikTok videos into their portfolio when applying for jobs. P16, a church minister, told us:

\begin{quote}\textit{Now (applying for a job), they want to check your social media. So (it'll be beneficial) if I can put out some quality material on TikTok." [P16]}\end{quote}


P16 created numerous videos focused on biblical narratives. He viewed these videos, which highlighted his adept storytelling skills, as a part of his resume. Instead of showcasing his storytelling abilities in a real-life scenario, he presented and recorded them in TikTok videos, making it easy for potential employers to review and assess his proficiency.

In addition to proficiency practicing, participants also reported leveraging TikTok for business promotion. Some of them utilized TikTok to reach a wider audience. P18, a fitness instructor and show host who ran a special program for blind individuals on YouTube, wanted to raise awareness of his show amongst the blind community. Consequently, he also shared his show on TikTok:


\begin{quote}\textit{Our videos on TikTok are shorter, whereas on YouTube we got 10-minute or hour-long videos...I find it's really easy to find people who are visually impaired since I can use hashtags like "BlindTok", and I think they will find me and our show faster, too...And it seems the short videos tend to get more attention because a lot of people don't have a lot of time. So that's just a strategy. [P18]}\end{quote}


While the participants mentioned above indicated that TikTok was easier than YouTube in terms of content creation, P18 introduced an additional point. He explained that videos on TikTok could easily connect with a broader audience if the creator knew how to capitalize on the platform's algorithmic features. This underscored the unique impact of the TikTok platform's style. The use of hashtags also underscores the application of folk theory when blind users aim to promote their content, a topic that has been explored in previous work \cite{Devito2018,devito2022transfeminine}. 


\subsubsection{\textbf{TikTok for Community Engagement}}

As we've discussed in the previous two sections, participants used TikTok for daily entertainment and professional development; these two themes concern blind users' utilization of TikTok. In this section, we spotlight the interactions between blind users at a community level. As noted earlier, TikTok boasts a powerful algorithm capable of uniting blind users. Specifically, blind creators employed a hashtag dubbed "BlindTok" to connect with and attract other blind viewers. By engaging with the content produced by blind creators, audiences also experienced a strong sense of relatedness.



Therefore, blind users established their own "BlindTok" community via the hashtag. This formed a significant portion of blind users' activities on TikTok. All participants (N=30) relayed their social experiences within the BlindTok community, as well as interactions with people from other minority groups. Participants reported both receiving and extending support to other BlindTokers. P12, a member of a group of male BlindTokers, shared how the multifaceted features of TikTok facilitated diverse modes of communication, socialization, and mutual support among group members:


\begin{quote}\textit{...we came together to...spread the word about blindness, amputation, diabetes, dialysis, and kidney failure. We did a couple of duets dancing together. We stitched each other videos asking questions... We go live-streaming every Monday and Thursday...with a group of men who are my brothers and then put out information to the world about blindness. I feel accomplished. [P12]}\end{quote}



The formation of the BlindTok community was not confined to blind individuals alone but also involved other minority groups. According to some participants, in addition to being blind, they also identified with other minority groups (such as LGBTQ, deaf, etc). For instance, both P21 and P30 were blind and identified as non-binary. The intersection of multiple minority identities prompted some blind users to recognize that, apart from fostering unity within the BlindTok community, they also needed to advocate for the well-being of other minority groups. Some participants recounted how they cared for other minorities, as P20 shared:

\begin{quote}\textit{Some blindness, like Retinitis Pigmentosa, is a disease that does both the ears and the eyes. So that's why (some blind users) they're pushing for captions for the deaf community. One of my friend (both blind and deaf), I make sure all the videos I send to her have captions. There's been a few and she message me back, saying "no caption." I'm like, "crap, I'll tell you what's going on." So she can tell what they're saying. [P20]}\end{quote}


P20 demonstrated her caring for individuals with hearing impairments by ensuring her content was accessible to them. She also highlighted specific features, like captions, on TikTok that enhanced accessibility. Recognizing the importance of supporting other minorities, some participants disclosed that they offered assistance to people with various disabilities via TikTok, especially during the pandemic. P06 shared with us:

\begin{quote}\textit{There was one (video) I did about teaching disabled people applying for benefits...not just visual impairment and blindness...Then (I) also (helped) people apply for blue badges, which means you can use disabled parking...[P06]}\end{quote}


P06's narrative resonates with the challenging circumstances many individuals with disabilities faced during the pandemic. The world grew significantly less accessible during the pandemic, and individuals with disabilities were either unable or reluctant to seek assistance to address the accessibility issues. Consequently, creating videos on TikTok became a means to empower them to find solutions by themselves. Notably, by learning from these videos, they could maintain their anonymity without revealing their struggles ([P06]). Beyond offering help, some individuals chose to champion the rights of other minority groups. For example, P30 stated:

\begin{quote}\textit{TikTok is for all demographics...the queer community and the disabled community, I've made quite a few friends...who make similar content to me. I also use the platform to educate...I have been an advocate for minority rights, I study and talk about issues and rights in queer and disabled communities. [P30]}\end{quote}


Similar to previous participants, P30 drew comparisons between TikTok and other social media platforms like Facebook and Snapchat. They acknowledged that TikTok, being one of the most popular social media platforms globally, had a far larger and more diverse user base than other platforms. This advantage enabled P30 to locate communities they identified with and advocate on their behalf. The support provided by blind individuals was also reciprocated, with participants noting an increase in TikTok videos where people spoke out against ableism, parallel to their stand against racism and sexism ([P09]).

\subsubsection{\textbf{Summary}} In this subsection, we detailed how TikTok's vlogging style (e.g., "talking head"), content features (e.g., videos with audio narratives), intuitive interactions (e.g., scrolling for browsing), and broad and diverse user groups (e.g., the BlindTok community) supported the myriad experiences of blind users. Collectively, these elements formed a TikTok infrastructure that supported the experiences of BlindTokers. These experiences also reflected blind individuals' perceptions of independence while using TikTok. The platform's casual style of content creation eased their performance-related anxieties; they leveraged TikTok as a means to hone their abilities towards leading more independent lives; they also stayed connected with other blind users via TikTok to mutually support each other's independence.


\subsection{TikTok's Accessibility Breakdowns and BlindTokers' Infrastructuring Work}


This subsection addresses the second research question. In the last subsection, we discussed how BlindTokers appreciated the TikTok platform and utilized it as an infrastructure for various purposes. However, the participants also highlighted numerous accessibility issues. These mainly pertained to visual content that was not accessible to blind users and compatibility problems where certain accessibility tools did not function effectively on TikTok. This led to accessibility breakdowns on TikTok, significantly impacting the experiences of BlindTokers. To illustrate the influence of accessibility breakdowns on BlindTokTokers' experiences, P15 said:

\begin{quote}\textit{And when you have all the filters at the bottom, I don't understand what any of those little icons mean. So I kinda get hung up on that filter. [P15]}\end{quote}

In the prior section, P15 informed us that he utilized one filter to articulate his visual impairment to his audience. However, this did not mean his content creation solely focused on showcasing his visual impairment. TikTok offers a variety of filters for creative content creation. Unfortunately, the small buttons on TikTok present accessibility challenges for blind users when it comes to understanding and selecting filters. This finding aligns with previous work where blind users expressed grievances about inaccessible visual filters \cite{Rong2022}. Ultimately, P15 found himself constrained to using only one filter consistently, which restricted his creative practices on TikTok.



In this section, we demonstrate more accessibility breakdowns and the infrastructuring work undertaken by participants to address these challenges with specific examples. Specifically, participants engaged in two types of infrastructuring work: the first involved acquiring and applying technical skills to find solutions, while the second centered around leveraging social relationships to overcome accessibility barriers.

\subsubsection{\textbf{Infrastructuring With Technical Skills}}


In this section, we present the technical practices employed by participants (N=25) to tackle accessibility breakdowns within the TikTok infrastructure. By "technical," we refer to the specific designs, features, tools, or devices that blind individuals utilized to navigate around infrastructure limitations. As previously mentioned, participants identified the small buttons as the most frustrating accessibility issue. To address this, some participants relied on their familiarity with the app. They memorized the button locations through "trial and error [P10]" and routinized the practices when using TikTok. However, TikTok frequently updated its interface, resulting in changes to the button layout. Blind users expressed dissatisfaction with these updates, as the updates disrupted blind users' established routines. P19 expressed their frustration, stating:

\begin{quote}\textit{I just recently realized when the bookmark feature got added, I was like, "wait, why is the clicking on the content creators page button in a different place", you get muscle memory from it. [P19]}\end{quote}


The "muscle memory" denoted how P19 developed a routine for using TikTok. However, the updates introduced by TikTok disrupted this infrastructure. As a result, blind users were forced to continually relearn the app's interface through laborious efforts. It is not uncommon to encounter accessibility issues resulting from system updates, as documented in prior research \cite{10.1145/3359293}. Faced with such breakdowns, the majority of participants reported attempting to utilize screen readers as a solution. Screen readers are assistive software programs that convert visual information into speech. However, participants often found that screen readers did not function effectively on TikTok. For instance, P14 shared:

\begin{quote}\textit{...when I get direct messages, I know it’s my inbox, because I use it, but (screen reader) it doesn't say “inbox”, it says like "4QLP658" [P14]}\end{quote} 


During the app development process, it is crucial for designers and programmers to ensure accessibility by providing clear descriptions for buttons, and allowing screen readers to read them aloud to blind users. However, P14's account revealed that the TikTok designer overlooked this aspect and used random numbers or letters instead. This is an ongoing issue that has been repeatedly highlighted by researchers in the field of accessible computing \cite{Ross2018}. In addition to the problem of non-labeled buttons, some participants experienced conflicts between certain functions of VoiceOver (screen reader on iPhone) and TikTok itself. P12 said:

\begin{quote}\textit{VoiceOver talks every now and then. It won't allow you to double-tap on certain things on TikTok. So sometimes you gonna turn it off...and kind of use a vision that I had to get where I need to be. But when I wanna "follow" someone, and then I need to turn VoiceOver on, sometimes when I turn VoiceOver back on, it'll jump back up to the previous videos. I got a swipe over again to get somewhere I was before. [P12]}\end{quote} 


In the first theme, blind users praised TikTok's interactions (double tapping) and content features (audio narratives). However, when users activated screen readers like VoiceOver, the experience turned unpleasant. Double-tapping the screen would trigger both TikTok and VoiceOver simultaneously, resulting in VoiceOver reading the screen information and creating interference while users were trying to consume the audio content from TikTok videos. Furthermore, the infrastructuring work of managing the incompatible interaction between the two apps (turning VoiceOver on and off) was not always successful and often led to additional unpleasant experiences. This finding aligns with previous research that highlighted blind individuals' reluctance to rely on assistive technologies due to concerns about potential breakdowns \cite{10.1145/3449223}.


Participants, cognizant of these issues, also engaged in comparisons of various assistive tools and opted for more reliable alternatives. Some participants chose external tools over the screen reader function on cell phones. P11, frustrated by the VoiceOver problems, decided to explore alternative options:

\begin{quote}\textit{But I don't know any of the things (buttons)... So I have a CCTV (magnifier) and I put my phone under that, and I was able to kind of see where on the screen a little bit better. [P11]}\end{quote} 


Many participants expressed a preference for external and physical tools after experiencing frustrations with VoiceOver. Some utilized physical magnifiers to read the small buttons ([P11]), while others employed external physical keyboards for video editing purposes ([P17]). However, it is important to note that arranging and incorporating external tools also required significant effort. For instance, P16, the church minister, illustrated how he created videos on Bible stories with multiple tools:


\begin{quote}\textit{(To make videos) I use green screen, selfie stick, tripod, and a remote control. When I wind up the camera, I'll put a piece of paper down to stand on. Then I go on, watch it, adjust until it's right. So sort of trial and error...my brother is a principal of a school, earns about \$80,000 per year, has a car, has a house, but when his kids see me, "wow, uncle, you have 1500 followers on TikTok!" [P16]}\end{quote}


The integration of different devices exhibited P16's infrastructuring work in addressing accessibility breakdowns. Despite the laborious process involved, his ability to overcome these hurdles actually enhanced his sense of independence. By successfully accomplishing these complex tasks, P16 experienced a heightened sense of self-efficacy, as evidenced by his nephews' comments. This aligns with the concept of self-efficacy proposed by Dixon et al. \cite{Dixon2007}.


As outlined above, the infrastructuring work involving technical skills typically occurred when blind users endeavored to address problems independently. In certain ways, TikTok played a role in motivating blind people to find solutions to accessibility issues on their own. P2 shared their perspective on this matter:

\begin{quote}\textit{...asking for help takes away some of that independence. If I do a presentation at school, I will usually ask for help...because I'm getting graded on it. But TikTok, I would just like to be able to do it independently. [P02]}\end{quote}

According to P02, there were different contexts for accessibility problems. Accessibility problems should receive greater attention in work settings due to the potentially serious consequences they can entail. Conversely, the accessibility issues encountered on TikTok were deemed less severe, likely due to the platform's playful nature and its peripheral role in participants' lives. The lower priority assigned to TikTok allowed participants to seek less assistance and, consequently, fostered a sense of independence while using the platform. This reflects participants' desire to engage in activities without relying on external help, highlighting their aspiration for self-reliance \cite{REINDAL1999}.


\subsubsection{\textbf{Infrastructuring Through Social Interactions}}



In addition to technical skills, participants (N=27) engaged in infrastructuring work through social interactions. While they made significant efforts to address accessibility breakdowns, there were certain challenges that proved too difficult to tackle on their own. Consequently, participants opted to confront and resolve these issues by engaging with individuals within their social networks. In most cases, participants sought to address problems alongside other BlindTokers. As previously mentioned, blind users shared aspects of their daily lives and provided techniques for living with visual impairments (such as learning Braille with dice) within the BlindTok community. Participants preferred collaborating with other BlindTokers for infrastructuring work because the support provided by the BlindTok community was more efficient and sustainable. P10 shared with us:

\begin{quote}\textit{If I do need help, I'll ask in the Blind (Tok) community...Because some sighted people, they don't know what the heck you're talking about...a lot of it has to do with using a screen-reader. I had no idea how to take a volume control for the music...and (a BlindToker) she showed the audience how to do it. But no sighted person would know how to do that unless they were trained in the assistant technology world. And I always try to be careful to describe what's going on in my videos. I had a picture of my birthday cake. And I explained this is an apple cake that a friend made me. [P10]}\end{quote}


Indeed, the concept of "infrastructure" itself is inherently relational \cite{Star1996}. Infrastructure breakdowns manifest in diverse ways, contingent upon the unique circumstances of each participant. The situated nature of infrastructuring work also renders it highly personal and context-dependent. As P10 rightly expressed, when confronted with accessibility breakdowns, learning from fellow blind users proved more beneficial than seeking assistance from sighted individuals. Blind users possess a deeper understanding of the situation and can offer more pertinent advice. Through such social interactions, P10 not only received support from the BlindTok community but also raised awareness about the importance of making her videos accessible to aid other blind users. Occasionally, participants would also reach out to sighted strangers for assistance, as mentioned by P02:


\begin{quote}\textit{...sometimes when you're watching videos on TikTok, people will say, look at this, but then they won't describe what this is. I will sometimes leave a comment on there like, hey, I'm visually impaired. I can't see that, what is that? And a lot of times people are very friendly and get back to you on that. They'll respond to you and tell you what it is. [P02]}\end{quote} 

P02's strategy showed that, blind people were able to overcome the accessibility breakdowns with the commenting functions. Specifically, the "crowd-sourcing" approach also foregrounded the importance of social factors, i.e. the blind-friendly atmosphere, of TikTok, when addressing accessibility problems. This echoes previous work on getting remote assistance from sighted strangers from online sources \cite{Lee2020,Brady2013}.

P02's approach demonstrated that blind individuals could effectively navigate accessibility breakdowns with the commenting function. Their strategy, akin to a "crowd-sourcing" approach, highlights the significance of social factors, particularly sighted TikTokers' awareness of making TikTok a more accessible place and being ready to help blind people when needed. This finding aligns with previous research on seeking remote assistance from sighted strangers through online platforms \cite{Lee2020,Brady2013}.


In addition to engaging with strangers, both blind and sighted, participants also tapped into their social relationships with friends to address accessibility breakdowns. P13, a financial consultant with aspirations of becoming an influencer on TikTok, shared her strategies for navigating these challenges:

\begin{quote}\textit{I volunteer a lot for these delta gamma girls in local colleges. So I just email one "can you come over and help me?"...I have a handful of people that I would call on...You can see the writing and cover things on my videos are pink. That's all me, they know I like pink and emojis. The video comes out with my flavor. [P13]}\end{quote}

P13's approach to infrastructuring work differed from the previously mentioned participants. While others sought instructions or information from others, P13 chose to have others do the work for her. However, P13 deliberately selected individuals whom she had previously assisted when she needed help. P13's story indicated how she orchestrated the video creation: the assistance in terms of making videos came from reciprocity \cite{Breheny2009} ("I volunteer a lot") and she had autonomy \cite{Hillcoat-Nalletamby2014} while creating TikTok videos. The story also underscores the importance of interdependence between blind users and helpers, as they work together to foster independence \cite{Bennett2018}.



In this section, the focus has predominantly been on how blind individuals receive support from their social networks. Besides, participants also reflected on their perceptions of independent living when they found themselves needing assistance from others. As per the perspective shared by many participants, seeking help as a visually impaired person did not contradict the notion of independence since everyone requires assistance at times ([P07]). It was the way of getting help and who to get help that affected the perceptions of independence. We will unpack participants' nuanced perceptions of independence in the discussion section.



\subsubsection{\textbf{Summary}} To summarize, blind users encountered numerous accessibility problems when using the TikTok infrastructure. Participants employed various strategies to address these problems, including technical skills (e.g. memorizing buttons, using smartphone apps, or using external tools) and social support (e.g. getting help from BlindTokers, sighted strangers, and friends). These strategies shed light on how blind individuals perceive and strive for independence when faced with accessibility issues. Generally, blind users preferred to solve problems independently, either by themselves or with the aid of reliable tools. If assistance was necessary, they often preferred seeking help from blind individuals or sighted strangers rather than relying on their family or caregivers. The discussion section will further explore how TikTok mediates these perceptions of independence and delve into the nuanced understanding of perceived independence.


\section{Discussion}




In the findings, we presented various activities that blind users had on TikTok. Such activities were mediated by different TikTok designs, such as browsing and sharing content, creating videos, and interacting with other users. The various activities also showed that participants considered TikTok and associated experiences as a foundation for their conception and construction of independence; therefore, we can also view TikTok as an infrastructure. Moreover, the activities also presented BlindTokers' understandings of TikTok: 1) blind users used TikTok in a casual way so they needed less help; 2) participants strove for professional success as well as independent living through TikTok; 3) BlindTokers stayed united with each other, as well as other minority groups, to help them be independent; 4) when encountering accessibility problems, BlindTokers' preferred less dependent solutions. Therefore, these understandings of TikTok also showed their perceptions of independence. We call the perceived independence of BlindTokers "\textit{TikTok-mediated independence}." In this section, we delve into TikTok-mediated independence by unpacking it from a multilevel perspective. We also critically discuss the infrastructuring work carried out by blind users to build such independence.

\subsection{Delving into the TikTok-mediated Independence With A Multilevel Perspective}

We introduce \textit{TikTok-mediated independence} as the most important theoretical contribution. TikTok-mediated independence is BlindTokers' \textit{perception} of independence when using TikTok; it represents BlindTokers' inner desire for independent living, and it is also mediated by TikTok's socio-technical features. TikTok-mediated independence has four facets: 

\begin{enumerate}
  \item Compared to other ICTs, the peripheral role of TikTok gives BlindTokers less pressure when using it, and therefore makes them \textit{feel} more independent. TikTok is a platform primarily for fun; and compared to other video platforms like YouTube, TikTok's short-video and "talking head" style generally requires less effort in content consumption and creation. This echoes Hong's work on the use social media to reduce the feeling of being dependent \cite{10.1145/2145204.2145300}, and also emphasizes that a lower expectation on tasks helps improve independence.
  \item Compared to other social network sites with fewer blind users or less blind-friendly content, TikTok has a large population of blind users and more blind-friendly content. Therefore, BlindTokers could \textit{witness} and \textit{get inspired by} other blind people's independent living. Recently, some apps like remote sighted services \cite{10.1145/3449223} allow blind people to get help from a large population of online users; TikTok also provides this, but with a more specific population that has similar backgrounds with blind people.
  \item TikTok's vlogging style encourages BlindTokers to \textit{present} their independent lives to their audience. The independence can be improved by the ease of using TikTok to shoot a video (compared to YouTube \cite{Seo2021}); but it can also be harmed by inaccessible TikTok designs.
  \item The powerful algorithmic features \cite{Devito2018} of TikTok help BlindTokers \textit{evaluate} the meaning of TikTok videos related to independence construction. That is to say, BlindTokers would feel more successful and independent when their videos get good algorithmic metrics.
\end{enumerate}

The four aspects are deeply connected to TikTok's socio-technical features and are mutually informed by each other. Although some of the aspects can also be found in other technologies, together they form a unique type of perceived independence among blind users.

\begin{figure}[!tb]
\centering
\includegraphics[width=0.45\linewidth]{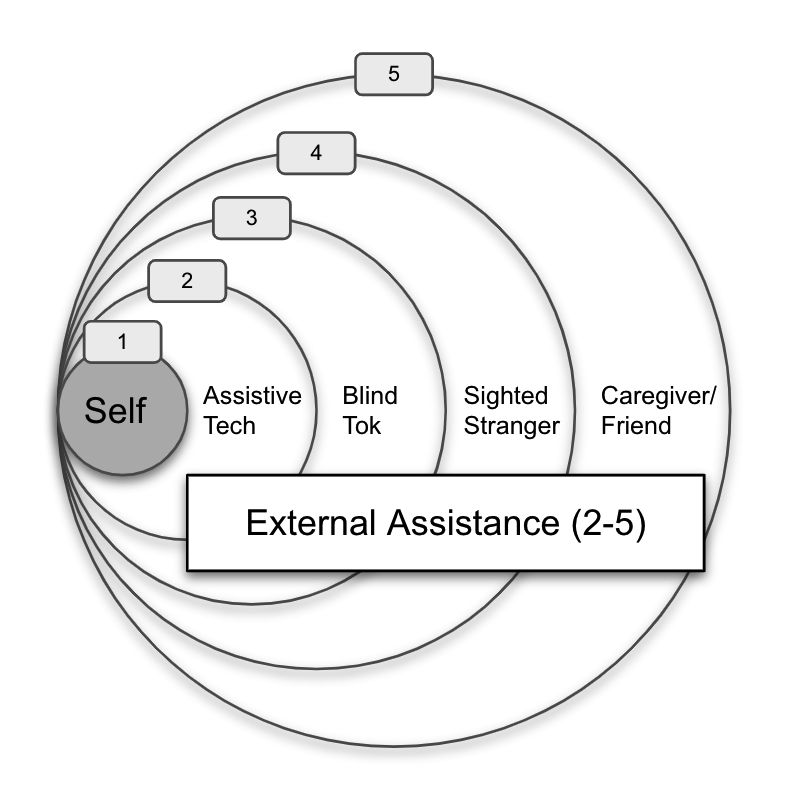}
\vspace{-0.1cm}
\caption{The Different Levels of TikTok-mediated Independence. }
\label{discussion}
\end{figure}

Based on the findings, we also illustrate the multilevel of TikTok-mediated independence with Figure ~\ref{discussion}. The multilevel perspective shows that blind users are aware that it is impossible to not ask for help while using TikTok. But some approaches are relatively more independent (as presented in the figure, self>assistive technology>BlindTok>sighted stranger>caregiver/friend). The first level is to complete tasks without any assistance; it usually happens when assistive technologies break down and the tasks are not so important. The second level shows that tasks are completed only with assistive technology (especially physical ones), such as magnifiers and physical keyboards. The third level is about asking for help from the BlindTok community. BlindTokers are more familiar with the difficulties, and participants can learn how to do tasks so that they would not need to ask others for help later. The fourth level is to seek the help of strangers on TikTok who are mostly sighted. On TikTok, participants can just leave a comment and wait for responses from whoever is available and willing to help; it helps participants ask less of their family/friends. The last level talks about getting help from people who have previously been helped by the participant. 






Our discoveries could help future studies better understand the nuance of independence. Researchers have been unpacking the relational nature of independence \cite{10.1145/3274354,10.1145/3449223}. In a most recent study on independence, Lee et al. \cite{10.1145/3449223} found that independence is largely influenced by blind users' situated perceptions: "\textit{Why, if the white cane is a tool designed to assist visually impaired people, do some PVI not think of it in the same way as other assistive technologies that might make them feel less independent? Where is the boundary between what is considered a skill or an extension of the self and what is considered an external tool that could detract from one’s independence?} (p.14) \cite{10.1145/3449223}" The questions of the "boundary" show that blind users think independence "\textit{not as a static or slow-changing personality trait} (p.6) \cite{10.1145/3449223}," they actually construct their perception of independence contextually. 

These questions also apply to the multilevel TikTok-mediated independence, where participants preferred some approaches over others. Therefore, we answer the questions by pointing out the key factors that affect blind users' preferences at each level: 

\begin{enumerate}
  \item \textbf{Capability of self}. For example, participants prefer to memorize buttons when it is impossible to read them with screen readers. This echoes the self-reliance \cite{REINDAL1999,10.1145/3449223} dimension of independence.
  \item \textbf{Reliability of tools}. Compared to electronic devices (which usually have complex mechanisms), blind users prefer simple and easy technologies that can be physically touched and reliably controlled \cite{Wehmeyer1996}, so that they do not need to worry about potential breakdown.
  \item \textbf{Similarity of background}. Blind people prefer to ask blind users, who are more familiar with living with visual impairments and using assistive technologies, for better instructions. Blind users can also learn from the instructions so that they no longer need assistance in the future, which contributes to self-efficacy \cite{Dixon2007}.
  \item \textbf{Availability of assistance}. Blind people tend to ask for help when there are abundant (preferably different) people who can provide assistance, because asking the same people multiple times would make blind people feel like they are bothering others.
  \item \textbf{Reciprocity when receiving assistance}. When asking friends/caregivers for help, blind users would like to choose people who have received their help. The freedom on selecting people to get assistance and the reciprocal relationship make blind people feel less dependent. This presents the influence of reciprocity \cite{Breheny2009}, interdependence \cite{Bennett2018}, and autonomy \cite{Hillcoat-Nalletamby2014}.
\end{enumerate}


The multilevel perspective not only shows the connection to previously discovered dimensions of independence but also situates them in specific contexts where blind users prioritize one over another.

The findings of this study also provide an actionable explanation for the roles of independence and interdependence in terms of evaluating accessibility. Accessible computing scholars have been discussing independence and interdependence in many ways \cite{Bennett2018}. One of the most important points, which is also closely relevant to our study, is that both independence and interdependence scholars are interested in the relationship between users with disabilities and external assistance. But how to differentiate which "relation" is for independence and which is for independence? For instance, based on our findings (as well as previous studies \cite{10.1145/3274354,10.1145/3449223,10.1145/1978942.1979424}), it is not uncommon to see people navigate the "relation" actually for independence (which can also be understood as interdependence). We provide an explanation: it is not about the focus on "relation" but the roles that differentiate independence and interdependence. For example, independence and interdependence both appear in our findings, but independence is the perceived goal while interdependence is one of many approaches, just like autonomy, control, and self-reliance, to the goal. Like Bennett acknowledged, "\textit{to achieve the goal of independent living, activists relied significantly on interdependent relationships with one another}. (p.163) \cite{Bennett2018}" The goal usually is perceived and evaluated by the users, while the approach can be analyzed by the designers and developers. This explanation is consistent with the essence of TikTok-mediated independence, which emphasizes BlindTokers' perception of independence. Differentiating the roles of independence and interdependence provides a better way to evaluate accessibility: from the perspective of users with disabilities, perceived independence would be the primary indicator; for designers of assistive technology, system, or infrastructure, the degree of interdependence would be a better criterion.

\subsection{Infrastructuring Independence: A Critical Examination of the TikTok Infrastructure}





In the current study, we introduced a case in which users conducted various activities on TikTok. In this sense, TikTok can be considered the infrastructure for such activities. However, blind users reported having encountered a wide variety of disruptions caused by TikTok's inaccessible designs. Therefore, the infrastructure broke down because it failed to support blind users' activities. Infrastructure failures occur for many reasons, and one of them is exclusion. As Star pointed out, infrastructure is learned by being a member of it \cite{Star1996}; for some populations, being excluded means that they would not be able to access and get support from the infrastructure. In fact, the literature on infrastructure has documented how infrastructure fails to support certain populations due to exclusion problems \cite{Shen2020,10.1145/2858036.2858109}. With the findings of the current study, we specified such infrastructure breakdown due to exclusion in an accessibility context. The designers of TikTok excluded blind users, or at least did not pay equal attention to blind and sighted users, when designing the app. The exclusion caused by inaccessibility further brought on blind users various social and technical problems, and they had to solve the problems by performing infrastructuring work. Based on the findings, the infrastructuring work involves various techniques and interactions that arranged different socio-technical resources that were internal or external to TikTok. In a recent study, Simpson et al. \cite{simpson2023hey} reported a group of neurodiverse TikTokers' infrastructuring work to counter accessibility issues. They highlighted the importance of \textit{creative labor}, i.e., "\textit{the time, effort, and creativity needed to produce an artistic output} (p.7) \cite{simpson2023hey}." Our study also describes the creative labor of BlindTokers and further enriches the discussion of creative labor. For instance, BlindTokers used creative labor not only for more accessible videos but also for presenting the inaccessible world from a blind person's perspective (e.g., using a filter to demonstrate visual impairment), which might add new approaches to understanding the contextual and situated nature of accessibility.

Our work extends the current literature on infrastructuring work by introducing infrastructuring work cases that occurred at the intersection of algorithmic exclusion and accessibility breakdown. Algorithmic exclusion is a new type of infrastructural exclusion; it manifests as "\textit{the ways in which algorithms construct and reconstruct exclusionary structures within a bounded sociotechnical system, or more broadly across societal structures.} (P.24)\cite{10.1145/3432951}" Semaan \cite{10.1145/3359175} has reported how marginalized people, like LGBTQ+ people, conducted infrastructuring work to overcome infrastructural exclusion. The work had many dimensions and three of them were closely relevant to our study: generating competence ("\textit{make sense of how infrastructures work and fail, which is a critical component in enabling human systems to develop new infrastructures}" (p.19) \cite{10.1145/3359175}), restoring control ("\textit{not relying on existing infrastructures and infrastructural arrangements}" (p.19) \cite{10.1145/3359175}), and revising normativity ("\textit{through infrastructuring practice... to build capacity to gain a foothold in their daily lives...[and] to counter what is normative}" (p.19) \cite{10.1145/3359175}). In our findings, we also find such dimensions of infrastructuring work. While Semaan's work showed that infrastructuring work contributed to resilience, i.e., the ability to bounce back from disruptions, our findings foreground how such work creates independence. BlindTokers made sense of TikTok by understanding and addressing its accessibility problems. They worked to be independent by relying less on human assistance or using external, reliable assistive technologies (e.g. magnifier, keyboard). To improve other BlindTokers' experiences, they also actively advocated for more accessible videos (highly audio-based) or made such videos by themselves, which countered the current fashion of TikTok (highly visual-oriented). By shedding light on BlindTokers' encounters with and resistance to algorithmic exclusion, we argue for more attention to algorithmic exclusions in accessibility contexts.






In addition, we also need to point out the potential ableism hidden behind the infrastructure of TikTok. Ableism is "\textit{a system that places value on people’s bodies and minds based on societally constructed ideas of normalcy, intelligence, excellence and productivity.} \cite{LEWIS}" As we introduced in the last section, TikTok shapes BlindTokers' perception of independence in many significant ways. One of them is about presenting independence to the audience and the quantified feedback on the presentation of independence, such as the number of likes, comments, views, and followers. This feature motivated BlindTokers to overcome the accessible issues by themselves and even made them feel fulfilled by their hard work. However, it could actually downplay the internal ableism of TikTok. Recently, infrastructure studies have paid growing attention to infrastructures that actually do not support but constrain certain populations' activities, such as surveillance infrastructure \cite{Li2021}. In our case, BlindTokers consumed many videos showcasing independence and therefore got inspired; the pervasiveness of inspirational videos might create a culture that justified the infrastructuring work to present independence. We continue the criticism of such infrastructure by shedding light on short-video platforms' potential infrastructural bias against blind people. Actually, there has been extensive research criticizing TikTok's oppression of populations like LGBTQ+ people \cite{rauchberg2022shadowbanned,10.1145/3432951}, Jewish \cite{divon2022jewishtiktok}, and people with disabilities \cite{bart2022true,simpson2023hey}. Previous studies also pointed out that blind users acutely noticed some productivity technology's inherent ableism \cite{Saha2020}. However, in our study, participants' reactions were not so strong. Probably because of the playful nature of TikTok, its underlying ableism is less likely to be identified and resisted. Research has shown that the playful side of technology could reshape or soften its negative and oppressive aspects, such as surveillance technologies \cite{Albrechtslund2008}. TikTok also has been portrayed as a playful platform among people with disabilities; it encourages users to create content that is "performative, exaggerated, and dramatized (p.12) \cite{10.1145/3411764.3445303}." By critically investigating the meaning of BlindTokers' infrastructuring work, we argue that the playful side of TikTok also has blurred the boundary between motivating self-representation and perpetuating ableism. In the future, more studies on critically investigating these platforms should be conducted, especially on foregrounding the potential ableism and drawing public attention to such issues.

\section {Design Implications}
Based on the inaccessible designs reported in the findings, we propose two categories of design implications for both TikTok and related assistive technologies. The first category deals with improving the compatibility of external accessibility functions and TikTok. According to the participants' descriptions, TikTok was not compatible with screen-readers on their cellphones: 1) many of TikTok's buttons are not labeled for screen-readers, so participants could not read them; 2) when using TikTok's comment function, participants could not dictate or review their comments; 3) some important interactions on TikTok conflict with operations of screen-readers, such as double tapping. Therefore, we recommend designers of both TikTok and screen-readers (e.g. VoiceOver) can collaborate with each other and resolve the issues by labeling all buttons, allowing dictation, and avoiding conflicting interactions. The second type of implication suggests a blind-friendly version of TikTok. Recently, many smartphone apps released special modes that are more accessible. For instance, WeChat released "\textit{Easy Mode} \cite{Tecent2021}" specifically for elder users; the easy mode features "\textit{bigger fonts, a clean layout, a text-to-speech feature, steer clear of any professional or internet lingo, and prohibits all forms of advertising and click-baits} \cite{PingWest2021}." Inspired by this, we also suggest a TikTok version specifically for blind users. Participants complained about 1) the prevalence of videos or pictures that have few audio descriptions, 2) the low contrast, small text fonts that are hard to read, and 3) frequent updates that change the display of buttons. Therefore, the blind-friendly version of TikTok could feature content with more audio descriptions, large and high-contrast text/buttons, and fewer updates (or updates that do not change the display of buttons).

\section {Limitations and Future Work}
There are two main limitations of this study. First, this project recruited participants through convenience sampling: due to the restriction of TikTok, we could only reach out to BlindTokers who followed our account. In addition, we only recruited English speakers. Therefore, although we had 30 participants coming from various backgrounds and we described BlindTokers' diverse behaviors, perceptions, and feelings with thick descriptions, we still acknowledge that our approach could cause generalizability issues. Furthermore, the data were collected in the summer of 2022. As reported by the participants, TikTok continuously updates its design; it could become more accessible or not in the future. Therefore, the findings of the current study only snapshot the state of TikTok in terms of accessibility by the summer of 2022. That said, this paper's contribution is more than evaluating the accessibility of the current version of TikTok; we also report how BlindTokers overcame unexpected accessible issues and how they socialized with others on TikTok, which are much more important to the literature on HCI and Accessible Computing.

Despite the limitations, this work lays solid ground for future studies on BlindTokers. We have identified that there is a huge community of blind users on TikTok, i.e. BlindTok. It is significantly bigger than blind communities on other social media like Facebook or Twitter. Considering the blind community on video-based platforms has been drawing growing attention recently \cite{Seo2021a}, future scholars can conduct more in-depth studies on the BlindTok community. In addition, some of the findings presented the collaboration among multiple users, including BlindTokers and sighted TikTokers. But in this study, we are more interested in the context where collaboration was triggered instead of the form or quality of collaboration. However, the popularity of collaboration among people with different ability statuses on TikTok also suggests that there could be new forms of crowd work supported by short-video platforms. Future studies can follow this direction to gain a more in-depth understanding of crowd work and interdependence among BlindTokers and sighted TikTokers. 


\section {Conclusion}
There has been extensive research on the experiences of people with visual impairments on social media platforms like Facebook and Twitter. However, little is known about the interactions between visually impaired people and short-video platforms like TikTok. We bridge the gap by presenting an interview study on 30 blind users on TikTok. We find three main types of activities that BlindTokers take on TikTok: everyday entertainment, community engagement, and professional development. By discussing the findings with prior work, we foreground BlindTokers' contextual understanding of independence as visually impaired people on TikTok; we also emphasized that independence, in BlindTokers' perceptions, is not merely a personality; it is an outcome that requires blind users' infrastructuring work. Based on the findings, we also provide design implications for a more accessible TikTok.

\begin{acks}
We sincerely thank the reviewers' comments that have greatly improved the quality of this paper. We also thank the assistance of Jiyoon Kim, Mark Anastasi, and He Albert Zhang in this project. We also appreciate all participants' contributions to this project.
\end{acks}

\bibliographystyle{ACM-Reference-Format}
\bibliography{PVI_TikTok}

\end{document}